\documentclass[preprint,10pt,3p]{elsarticle}
\usepackage{amsmath,amssymb,amsthm,graphicx,url}
\usepackage{mathtools} 

\usepackage{balance}
\usepackage{caption}
\usepackage{subcaption} 

\usepackage{booktabs}
\usepackage{multirow}
\usepackage{setspace}
\usepackage{times}

\usepackage{xspace}

\newcount\Comments  
\usepackage{color}
\newcommand{\kibitz}[2]{\ifnum\Comments=0\textcolor{#1}{#2}\fi}

\definecolor{olive}{rgb}{0.23,0.70,0.44}

\definecolor{dark_green}{rgb}{0,0.51,0.0}

\makeatletter
\def\ps@pprintTitle{%
   \let\@oddhead\@empty
   \let\@evenhead\@empty
   \def\@oddfoot{\reset@font\hfil\thepage\hfil}
   \let\@evenfoot\@oddfoot
}
\makeatother

\newcommand{\catvehicle}{CAT Vehicle\xspace}
\newcommand{\cv}{\catvehicle}

\usepackage{amssymb}

\journal{Transportation Research Part C}

\begin{document}

\begin{frontmatter}

\title{Dissipation of stop-and-go waves via \\control of autonomous vehicles: Field experiments}



\cortext[cor1]{Corresponding author}




\author[a]{Raphael E. Stern\fnref{fn1}}
\author[b]{Shumo Cui\fnref{fn1}} 
\author[c]{Maria Laura Delle Monache\fnref{fn1}}
\author[d]{Rahul Bhadani}
\author[d]{Matt Bunting}
\author[a]{Miles Churchill}
\author[e]{Nathaniel Hamilton}
\author[f]{R'mani Haulcy}
\author[g]{Hannah Pohlmann}
\author[a]{Fangyu Wu}
\author[h]{Benedetto Piccoli\fnref{fn2}}
\author[b]{Benjamin Seibold\fnref{fn2}}
\author[d]{Jonathan Sprinkle\fnref{fn2}}
\author[a]{Daniel B. Work\fnref{fn2}\corref{cor1}}
\ead{dbwork@illinois.edu}

\address[a]{Department of Civil and Environmental Engineering, University of Illinois at Urbana-Champaign, 205 N. Mathews Ave, Urbana, IL 61801, USA.}
\address[b]{Department of Mathematics, Temple University, 1805 North Broad Street, Philadelphia, PA 19122, USA.}
\address[c]{Inria, University Grenoble Alpes, CNRS, GIPSA-lab, F-38000 Grenoble, France.}
\address[d]{Electrical and Computer Engineering, University of Arizona, Tucson, AZ 85721-0104, USA.}
\address[e]{Lipscomb University, 1 University Park Drive, Nashville, TN 37204, USA.}
\address[f]{Yale University, New Haven, CT 06520, USA.}
\address[g]{Pennsylvania State University, University Park, PA 16801, USA.}
\address[h]{Department of Mathematical Sciences, Rutgers University -- Camden, 311 N. 5th St, Camden, NJ 08102, USA.}

\fntext[fn1]{R. Stern, S. Cui, and M.L. Delle Monache contributed equally to this work.}
\fntext[fn2]{B. Piccoli, B. Seibold, J. Sprinkle, and D. Work contributed equally to this work.}

\begin{abstract}
Traffic waves are phenomena that emerge when the vehicular density exceeds a critical threshold. Considering the presence of increasingly automated vehicles in the traffic stream, a number of research activities have focused on the influence of automated vehicles on the bulk traffic flow. In the present article, we demonstrate experimentally that intelligent control of an autonomous vehicle is able to dampen stop-and-go waves that can arise even in the absence of geometric or lane changing triggers. Precisely, our experiments on a circular track with more than 20 vehicles show that traffic waves emerge consistently, and that they can be dampened by controlling the velocity of a single vehicle in the flow. We compare metrics for velocity, braking events, and fuel economy across experiments. These experimental findings suggest a paradigm shift in traffic management: flow control will be possible via a few mobile actuators (less than 5\%) long before a majority of vehicles have autonomous capabilities. 
\end{abstract}

\begin{keyword}
Traffic waves, Autonomous vehicles, Traffic control
\end{keyword}

\end{frontmatter}

\section{Introduction}
\subsection{Motivation}

The dynamics of traffic flow include instabilities as density increases, where small perturbations amplify and grow into stop-and-go waves that travel backwards along the road~\cite{treiterer1974,Sugiyamaetal2008,FlynnKasimovNaveRosalesSeibold2009,kerner2012physics}. These so-called \textit{phantom} traffic jams are an experimentally reproducible phenomenon, as demonstrated in different experiments \cite{Sugiyamaetal2008,Tadakietal2013,jiang2014traffic}. 
Common wave triggers 
include lane changing \cite{laval2006lane, laval2006stochastic, zheng2011applications}, but they can even be generated in the absence of any lane changes, bottlenecks, merges, or changes in grade~\cite{Sugiyamaetal2008,Tadakietal2013}.  
Moreover, these waves can be captured in microscopic models of individual vehicle motion \cite{BandoHesebeNakayama1995,NagelSchreckenberg1992,Garavelloetal2016} (see also the reviews \cite{brackstone1999car,chowdhury2000statistical, Helbing2001}) and macroscopic models described via solutions to continuum problems \cite{FlynnKasimovNaveRosalesSeibold2009,Payne1971,Whitham1974,AwRascle2000,Zhang2002,Greenberg2004}. Since these waves emerge from the collective dynamics of the drivers on the road, they are in principle avoidable if one could affect the way people drive. Recognizing the rapid technological innovations in traffic state estimation and control, this work provides experimental evidence that these waves can be reduced by controlling a small number of vehicles in the traffic stream. 

A necessary precursor to dissipating traffic waves is to detect them in real-time. Advancements in traffic state estimation~\cite{gazis1971line,wang2005real,Onsequential} have facilitated high resolution traffic monitoring, through the advent of GPS smartphone sensors \cite{herrera2010evaluation,work2010traffic,de2008traffic,hofleitner2012learning} that are part of the flow---termed \textit{Lagrangian} or \textit{mobile} sensors. Now commercialized by several major navigation services, the use of a small number of GPS equipped vehicles in the traffic stream has dramatically changed how traffic is monitored for consumer-facing mobility services, which previously relied on predominantly fixed sensing infrastructure.

Currently, traffic control is dominated by control strategies that rely on actuators at fixed locations or are centralized. Such systems include \textit{variable speed advisory} (VSA) or \textit{variable speed limits} (VSL) \cite{nissan2011evaluation,hegyi2005optimal,  smulders1990control,hegyi2008specialist,popov2008distributed}, which are commonly implemented through signs on overhead gantries, and ramp metering~\cite{papageorgiou2000freeway,gomes2006optimal, papageorgiou1991alinea}, which relies on traffic signals on freeway entrance ramps. More recently, coordinated systems to integrate both ramp metering and variable speed limits have been proposed \cite{hegyi2005model,papamichail2008integrated,lu2010combining,han2017resolving}. A common challenge of VSL and ramp metering systems is the small flexibility of the systems due to the high cost of installation of the fixed infrastructure, which consequently limits the spatial resolution of the control input.

Recent advancements in vehicular automation and communication technologies provide new possibilities and opportunities for traffic control in which these smart vehicles act as Lagrangian actuators of the bulk traffic steam. When a series of adjacent vehicles on a roadway are connected and automated, it is possible to form dense platoons of vehicles which leave very small gaps. A key challenge for vehicle platoons is to design control laws in which the vehicle platoon remains stable, for which significant theoretical and practical progress has been made~\cite{levine1966optimal,swaroop1996string,shladover1995review,fenton1991automated,darbha1999, besselink2017string,ioannou1993intelligent,buehler2009darpa,rajamani1998design}.  In contrast to the vehicle platoon setting, in which all vehicles are controlled, or the variable speed limit and ramp metering strategies which actuate the flow at fixed locations, this research aims to dissipate congestion-based stop-and-go traffic waves using only a sparse number of autonomous vehicles already in the flow, without changing how the other, human-driven, vehicles operate. 

The notion to dissipate stop-and-go waves via controlling vehicles in the stream represents a shift from stationary to Lagrangian control, mirroring the transition to Lagrangian sensing that has already occurred. The key advantage in mobile sensing projects  \cite{herrera2010evaluation,de2008traffic,hofleitner2012learning} is that a very small number of vehicles being measured (3-5\%) suffices to estimate the traffic state on large road networks \cite{work2010traffic}. In the same spirit, our research experimentally demonstrates that a small number of Lagrangian controllers suffices to dampen traffic waves.

The ability of connected and automated vehicles to change the properties of the bulk traffic flow is already recognized in the transportation engineering community.  For example, the works~\cite{davis2004effect,talebpour2016influence,gueriau2016assess,wang2016cooperative} directly address the setting where a subset of the vehicles are equipped with automated and/or connected technologies, and then assess via a stability analysis or simulation the extent to which the total vehicular flow can be smoothed. Recently, several works have explored extensions to the variable speed limit control strategies in which connected or automated vehicles are used to actuate the traffic flow~\cite{wang2016connected}. For example, the work~\cite{Han2017113} develops a VSL strategy that is implemented in simulation with connected vehicles where the traffic evolves according to the kinematic wave theory. It follows a similar strategy proposed in~\cite{van2014cooperative}, where a coordinated VSL and ramp metering strategy is implemented via actuation of the entire vehicle fleet (i.e., 100\% penetration rate). Although not explicitly designed as a variable speed limit controller, the article~\cite{nishi2013theory} advocates a ``slow-in, fast-out'' driving strategy to eliminate traffic jams, using a microscopic model also in line with kinematic wave theory. The work~\cite{he2016jam} proposes a similar jam absorbing strategy as~\cite{nishi2013theory} based on Newell's car following theory, and its effectiveness is assessed in simulation.

Interestingly, an experimental test of the ``slow-in, fast-out'' strategy~\cite{nishi2013theory} is provided in~\cite{Taniguchi2015}, in which five vehicles are driven on a closed course. The lead vehicle in the platoon of five vehicles drives initially at a constant speed, then decelerates as if driving through a congestion wave, and then accelerates back to the cruising speed. The third vehicle in the platoon initially leaves a large gap, and due to the extra gap it is able to maintain the cruising speed and effectively absorb the jam. In contrast to the experiment~\cite{Taniguchi2015}, the present work fully replicates the setup of Sugiayama et al.~\cite{Sugiyamaetal2008,Tadakietal2013}, in which the stop-and-go wave is generated naturally from the human drivers in the experiment, without an external cause. Moreover, the controllers proposed in the present work are distinct. 

We also note some preliminary field experiments to harmonize speeds via connected vehicles are recently reported in~\cite{ma2016freeway,lu2015novel}, in part to measure the impact of connected vehicles following an infrastructure-generated advisory speed on the traffic stream behind the connected vehicles. In the present article, we instead dampen waves on a closed ring, which simplifies the experimental setup, and facilitates detailed data collection on the performance of the controllers.

\subsection{Problem statement and contributions }
 
The present article is inspired by the work of Sugiyama et al.~\cite{Sugiyamaetal2008,Tadakietal2013} which are the first works to demonstrate via experiments that traffic waves can emerge without infrastructure bottlenecks or lane changing. A series of experiments is conducted where approximately 20 vehicles drive in a ring of fixed radius with each driver following the vehicle in front of them. The experiments~\cite{Sugiyamaetal2008,Tadakietal2013} are foundational because they demonstrate the emergence of traffic waves caused (unintentionally) by human driving behavior. However, they do not offer a solution for dampening these waves.

To address this gap, we design and execute a series of ring-road experiments which show that an intelligently controlled autonomous vehicle is able to dampen stop-and-go waves. The experimental setup (described in Section~\ref{sec:experimental_methodology}) follows the setting of Sugiyama et al.~\cite{Sugiyamaetal2008,Tadakietal2013}, with the modification that one vehicle is an autonomous-capable vehicle which can run a variety of longitudinal control laws.   Similar to the Sugiyama et al.~\cite{Sugiyamaetal2008} experiment, the  position and velocity of each vehicle is tracked via a 360 degree camera. We additionally instrument each vehicle in the 22-car fleet (see Table~\ref{tab:vehicle_summary} in Appendix~\ref{sec:vehMake} for a detailed description of the vehicle fleet)  with an OBD-II scanner to log the real-time fuel consumption of each vehicle, such that the impact of the traffic waves and controllers on the bulk fuel consumption can be recorded. 

We present three experiments (labeled as A, B, and C in Section~\ref{sec:analysis_of_experimental_data}) and two distinct control strategies (detailed in Section~\ref{sec:controllers}) that can
be used to dampen stop-and-go waves created by human drivers. The first control strategy is to follow a fixed average velocity (selected based on observation) as closely as possible without collisions. It is implemented in Experiment A via an automatic control algorithm (called \emph{FollowerStopper}) and in Experiment B via a carefully trained human driver. The second type of control strategy is a \textit{proportional-integral} (PI) \emph{controller with saturation},
which is a natural extension of the PI controller, a simple and widely used controller in industrial applications.  The controller is only based on the knowledge of the autonomous vehicle speed over a time horizon. The control action is saturated at small gaps to avoid collisions, and long gaps to avoid slowing down of traffic. Compared to the average velocity controllers (Experiments A and B), the PI controller with saturation directly estimates the average velocity and thus needs no external input. 

The results of each of the three experiments are presented and compared in Section~\ref{sec:analysis_of_experimental_data}. In each experiment, stop-and-go waves arise dynamically when all vehicles are under human control. Once one vehicle is activated to be autonomous (with the control algorithms described in Section~\ref{sec:controllers}), the traffic waves are dissipated. Compared to when waves are present, the Lagrangian control results in up to 40\% less fuel consumption, and a throughput increase of up to 15\%. Future perspectives for Lagrangian vehicular control are provided in Section~\ref{sec:conclusions}.

\section{Experimental methodology} \label{sec:experimental_methodology}
We briefly describe the experimental setting in which stop-and-go waves are observed to develop and subsequently dampened via control of a single vehicle in the experiment (mimicking a low penetration rate on a long freeway stretch). The experiments follow the ring setting Sugiyama, et al.~\cite{Sugiyamaetal2008,Tadakietal2013}. A key advantage of the ring road experimental setup~\cite{Sugiyamaetal2008,Tadakietal2013} is that it removes other effects like boundary conditions, merging lanes, or intersections. To aid in interpretation of trajectory and fuel consumption datasets made available with this work, we concisely describe the experimental design and data collection methods in Section~\ref{sec:design}. The protocol for each experiment, including the specific instructions given to the drivers are presented in Section~\ref{sec:mechanics}. 

\subsection{Experiment design}\label{sec:design}

\begin{figure}
    \centering
    
        \begin{subfigure}[b]{\textwidth}
        \includegraphics[width=1.0\textwidth]{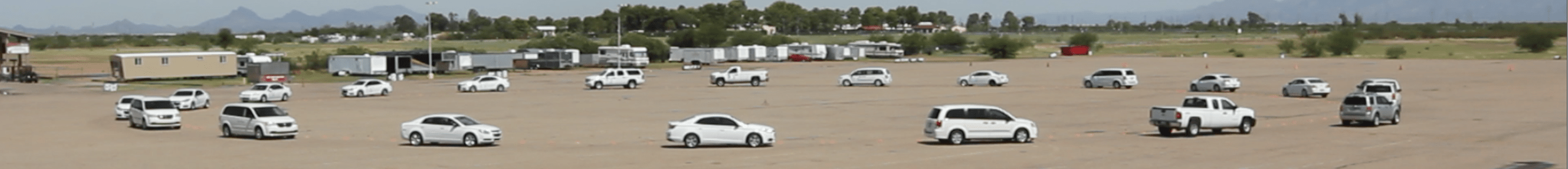}
        \caption{Alignment of vehicles at start of Experiment A.}
        \label{fig:experiment05Start}
        \end{subfigure}\\
        
        \begin{subfigure}[b]{\textwidth}
        \includegraphics[width=1.0\textwidth]{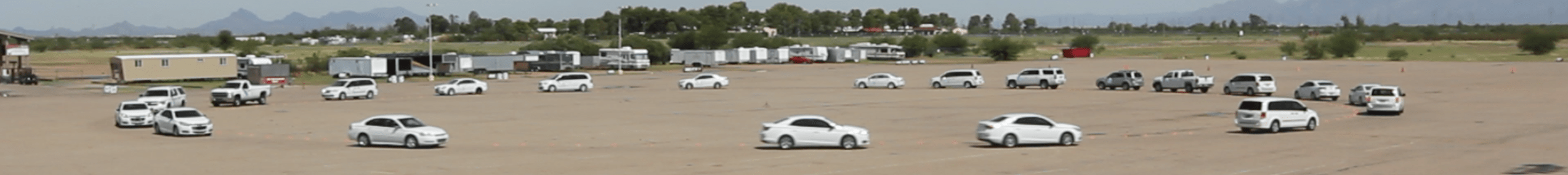}
        \caption{Alignment of vehicles 93 seconds into Experiment A when wave is present in back right.}
        \label{fig:experiment05StopNGo123s.}
        \end{subfigure}\\
        
        \begin{subfigure}[b]{\textwidth}
        \includegraphics[width=1.0\textwidth]{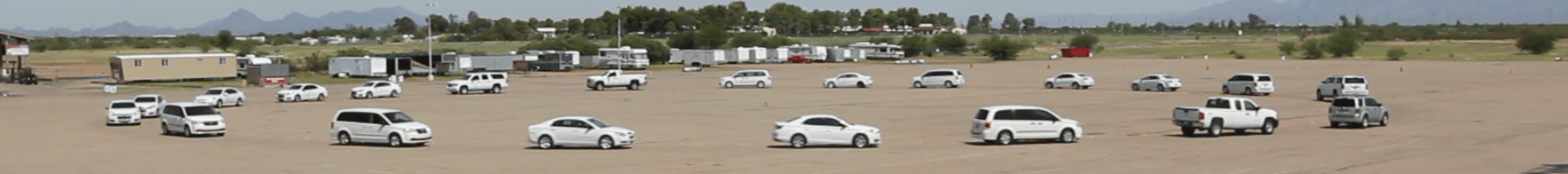}
        \caption{Alignment of vehicles 327 seconds into Experiment A when the \cv is actively dampening the wave.}
        \label{fig:experiment05StopNGo327s.}
        \end{subfigure}\\
        
    \caption{Ring track with 21 vehicles during Experiment A.
    }
    \label{fig:Experiment05Screenshot}
\end{figure}

We consider a single-lane circular track of radius 41.4 meters to the center of the lane (260 meter circumference) with 21 to 22 vehicles depending on the experiment (see Figure~\ref{fig:Experiment05Screenshot}). Small modifications to the Sugiyama et al.~\cite{Sugiyamaetal2008,Tadakietal2013} setup include a larger circumference of the circle and driving in counter-clockwise direction, to account for the larger average US vehicle size and the location of the steering wheel.  An asphalt track is marked with small circular cones, and is otherwise nearly flat and uniform (no marking, light poles, parking barriers, or other potential obstacles). Short (3 cm) orange indicators are placed to mark the inside of the track ring. 


A fleet of 22 passenger vehicles  
equipped with data acquisition hardware is used in the experiment. One of the 22 vehicles is the University of Arizona self-driving capable \textit{Cognitive and Autonomous Test} (CAT) Vehicle, which can be transitioned between manual velocity control and autonomous velocity control. A trained human driver controls the steering wheel of the \cv at all times during all experiments. We underscore that \emph{only one} vehicle is ever controlled to dampen the traffic wave, either via automation of the vehicle velocity or through the trained driver. This setting in which a single vehicle is controlled on a ring road mimics a low penetration rate of automated vehicles on a long stretch of highway. All other vehicles are driven by University of Arizona employees that have completed safe driver training but received no other special driving training. Drivers are instructed to drive safely and are requested to attempt to close any widening gaps between their vehicle and the vehicle ahead (see Section~\ref{sec:mechanics} for the precise driver instructions). 

Data from each experiment is collected via a video camera and OBD-II data loggers. The 360 degree camera is placed at the center of the track and used to record each experiment. The resulting video is processed via computer vision techniques to identify the center of each vehicle in each frame, which is then smoothed to generate the vehicle trajectories, following the approach described in~\cite{WuTRB2017}. Data from the in-vehicle devices are gathered through the OBD-II standard interfaces available on all US cars starting in 1996 \cite{obdii}. The in-vehicle devices measure the instantaneous fuel consumption of each vehicle. 

\subsection{Experiment mechanics} \label{sec:mechanics}
Each experiment lasts between 7--10 minutes to limit driver fatigue and begins with all vehicles uniformly spaced around the track according to the position of their front-left tire. The \cv begins each experiment in manual mode, and is switched into a control mode during the experiment. Traffic waves appear in all experiments, and the unsteady traffic is allowed to persist for at least 45 seconds before a controller is activated. For some controllers, a desired average velocity is communicated from an external observer.

Precisely, each experiment consists of the following phases: (i) setup; (ii) evacuation; (iii) initialize; (iv) drive; (v) stop; and (vi) conclusion, summarized below.

\begin{enumerate}
\item[i.] Setup: Vehicles are distributed equally according to the spacing of their front-left tire. Drivers are individually instructed to turn on their in-vehicle data recorders. Additional driver instructions (if any) are delivered to individual drivers through the window.
\item[ii.] Evacuation: The central camera is switched on. All research team personnel evacuate the track.
\item[iii.] Initialize: An air horn sounds to instruct all drivers to switch gears from Park to Drive, without moving.
\item[iv.] Drive: An air horn sounds, to instruct all drivers to begin driving.
\item[v.] Stop: An air horn sounds, instructing drivers to come to a safe stop and switch gears into Park.
\item[vi.] Conclusion: Experiment personnel enter the track after all vehicles have stopped. Drivers are individually instructed to turn off their in-vehicle recorders. The central camera is switched off.
\end{enumerate}

The following instructions are provided to each driver prior to the start of the experiments. ``Drive as if you were in rush hour traffic. Follow the vehicle ahead without falling behind. Do not pass the car ahead. Do not hit the car ahead. Drive safely at all times. Do not tailgate. But put an emphasis on catching up to the vehicle ahead, if a gap starts opening up.'' The purpose of these instructions is to explicitly prevent the human drivers from intentionally smoothing out traffic waves themselves. This is important, so that the wave-smoothing effect solely caused by the \cv can be studied. 
%

In the event of an unsafe scenario, the drivers are instructed to steer out of the circle, at which the experiment coordinator will sound the air horn and the experiment will stop. All drivers are instructed that the \cv would be switching back and forth between autonomous and manual mode, and that they should focus on their driving rather than attempting to guess what mode the vehicle is in at any given time. The vehicle directly following the \cv is told to drive as if the \cv were in the same lane, and that the \cv will be driving at a larger radius (1/2 vehicle width) 
to facilitate evasive emergency maneuvers.


\section{Description of controllers of the autonomous vehicle}\label{sec:controllers}
This section presents the velocity controllers implemented on the \catvehicle to dampen traffic waves on the ring track. The controllers are generally motivated by the fact that mathematical models of vehicular traffic can be stabilized via the control of a small number of vehicles, see for example~\cite{davis2004effect,talebpour2016influence,gueriau2016assess,wang2016cooperative,Cui2017}. One important notion that can help stabilize the overall flow is to have a subset of vehicles drive with a smooth driving profile relative to the traffic conditions, which is the basis of the works~\cite{van2014cooperative,nishi2013theory,he2016jam}. 

One possible way to create a smooth driving profile is to follow the average speed of the vehicles ahead, which drive faster then the average speed before a stop-and-go wave, and slower than the average speed during the wave. By simply driving at the average speed, a vehicle covers the same distance in the same amount of time, but with less acceleration and breaking. Additional logic is necessary to prevent extremely small gap (unsafe) situations from appearing, or large gaps that may induce lane changing on multi-lane roadways. With these ideas in mind, we propose two possible control laws and show experimentally that they are able to stabilize the flow on the ring.

The general structure of the controllers is as follows. The \catvehicle continuously tracks its velocity $v^\text{AV}$, and measures (at a sampling rate of 30Hz) the \emph{gap} $\Delta x$, defined as the distance from its front bumper to the rear bumper of its lead vehicle ahead. This signal, suitably smoothed, is used to calculate the \emph{velocity difference} $\Delta v = \frac{\text{d}}{\text{d}t}\Delta x$ between the lead vehicle and the AV. The lead vehicle's (i.e., the car ahead of the AV) velocity is estimated on-board the \catvehicle as $v^\text{lead} = v^\text{AV}+\Delta v$. Moreover, a \emph{desired velocity} $U$ is defined (obtained in various ways, see below), which, when chosen correctly, can stabilize the traffic flow. From the desired velocity, the gap, and the velocities of the \cv and lead vehicle, a \emph{commanded velocity} $v^\text{cmd}$ is determined.  The commanded velocity is then passed to a low-level controller on the \cv that translates it into an actuation of the accelerator or brake.  Note that all of these quantities are functions of time; but the time argument is frequently omitted for notational efficiency. Below, we describe: strategies to define a desired velocity, 
the \textit{FollowerStopper} controller and a PI controller with saturation, 
the low level controls, 
and a control law implemented by a trained human driver.


\subsection{The FollowerStopper controller}
\label{sec:followerStopper}


%

The premise of this controller is to command exactly the desired velocity $U$ whenever safe (i.e., as in a standard cruise controller), but to command a suitable lower velocity $v^\text{cmd} < U$ whenever safety requires, possibly based on the lead vehicle's velocity. 

\begin{figure}
    \centering
    \includegraphics[width = 0.6\textwidth]{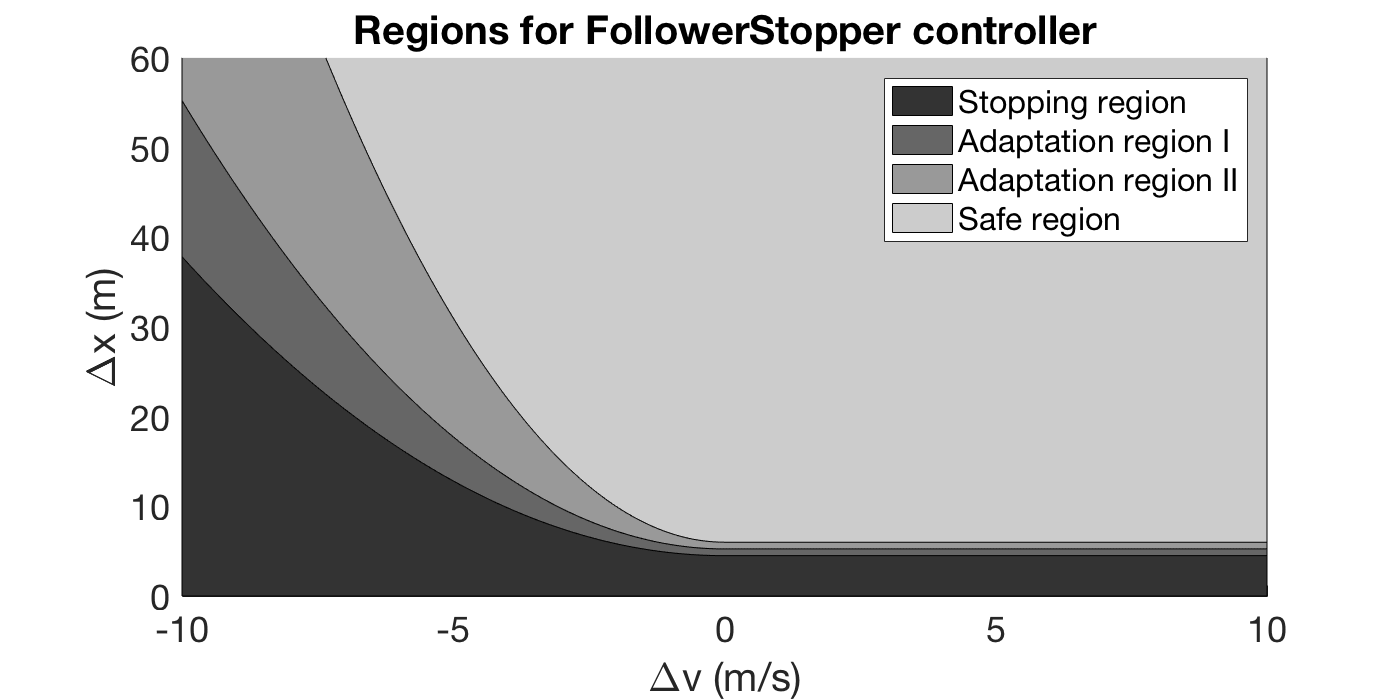}
    \caption{Regions defined in FollowerStopper controller.}
    \label{fig:followerStopper_regions}
\end{figure}

Using the gap $\Delta x$ and the velocity difference $\Delta v = \frac{d}{dt}\Delta x = v^\text{lead}-v^{AV}$, the $\Delta x$--$\Delta v$ phase space is divided into regions (see also Figure~\ref{fig:followerStopper_regions}):
\begin{enumerate}
\item[i.]a safe region, where $v^\text{cmd} = U$;
\item[ii.] a stopping region, where a zero velocity is commanded; and an 
\item[iii.] an adaptation 
region (two parts), where some average of desired and lead vehicle velocity is commanded. 
\end{enumerate}

The boundaries between the regions are parabolas in the $\Delta x$--$\Delta v$ phase space (trajectories that the AV/lead vehicle pair would traverse when decelerating at constant rates), defined as 
\begin{equation}
    \Delta x_k = \Delta x_k^0 + \frac{1}{2d_k}(\Delta v_-)^2\;,\quad \text{for } k=1,2,3\;.
    \label{eq:regions}
\end{equation}
Here $\Delta x_k^0$ is a parameter that defines the intercept in the $\Delta x$--$\Delta v$ phase space, and $d_k$ controls the curvature and is interpreted as a deceleration rate.  Moreover, $\Delta v_- = \min(\Delta v,0)$ is the negative arm of velocity difference, i.e., the case of the \catvehicle falling behind is treated just like the case $v^\text{AV} = v^\text{lead}$.

Using the region boundaries defined in~\eqref{eq:regions}, the commanded velocity is
\begin{equation}
v^\text{cmd} = 
\begin{cases}
0 & \text{if~~} \phantom{\Delta x_1 <\ } \Delta x \le \Delta x_1 \\
v \frac{\Delta x-\Delta x_1}{\Delta x_2-\Delta x_1}
& \text{if~~} \Delta x_1 < \Delta x \le \Delta x_2 \\
v+(U-v)\frac{\Delta x-\Delta x_2}{\Delta x_3-\Delta x_2}
& \text{if~~} \Delta x_2 < \Delta x \le \Delta x_3 \\
U & \text{if~~} \Delta x_3 < \Delta x\;.
\end{cases}\label{eq:fslogic}
\end{equation}
In \eqref{eq:fslogic}, the velocity $v = \min(\max(v^\text{lead},0),U)$ is the lead vehicle velocity (if positive) or the desired velocity, whichever is smaller. In the adaptation regions ($\Delta x_1 < \Delta x \le \Delta x_3$), the commanded velocity transitions continuously from stopping ($v^\text{cmd} = 0$, for short gaps) to safe driving ($v^\text{cmd} = U$, for large gaps), via a transition involving the lead vehicle's velocity.

 As implemented, in~\eqref{eq:regions} and \eqref{eq:fslogic} we set $\Delta x_1^0 = 4.5\text{ m}$, $\Delta x_2^0 = 5.25\text{ m}$, and $\Delta x_3^0 = 6.0\text{ m}$, and the deceleration rates are $d_1 = 1.5~\frac{\text{m}}{\text{s}^2}$, $d_2 = 1.0~\frac{\text{m}}{\text{s}^2}$, and $d_3 = 0.5~\frac{\text{m}}{\text{s}^2}$. Note that the $\Delta x_k$ boundaries of the regions depend strongly on the velocity difference between the \catvehicle and the lead vehicle. For instance, if the \catvehicle is catching up rapidly at $\Delta v = -3~\frac{\text{m}}{\text{s}}$, then $\Delta x_1 = 7.5\text{ m}$, $\Delta x_2 = 9.75\text{ m}$, and $\Delta x_3 = 15\text{ m}$.

\subsection{The PI with saturation controller}
\label{sec:PI_controller}
%
The idea behind this controller is that the \cv
may estimate the average speed of the vehicles in front, and then drive according to the average speed. When stop-and-go waves are present, it allows a gap to open up in front of the \cv when the lead vehicle accelerates, which is then closed when the lead vehicle decelerates. An estimate of the average speed required by the controller is obtained  by measuring
the \cv speed over a large enough time horizon.

The controller determines a command velocity $v^{\text{cmd}}$ following a standard proportional integral control logic~\cite{aastrom2008feedback}, where the
deviation from the average speed is treated as the error signal in the PI controller.
This simple idea needs to be paired with saturation:
for small gaps the \cv should follow the lead vehicle speed to avoid dangerous situations, while for large gaps, the \cv should catch up to the lead vehicle.

More precisely, this controller estimates the desired velocity, $U$, as a temporal average of the \cv's own velocity over an interval.  Letting $v^\text{AV}_1,\dots,v^\text{AV}_m$ denote the \catvehicle velocities over the last $m$ measurements, the desired velocity is computed as the temporal average $U = \frac{1}{m}\sum_{j=1}^m v^\text{AV}_{j}$. In practice, we choose $m$ corresponding to a 38 second interval, which is approximately the time required to travel one lap around the ring. 

The desired average velocity is then translated into a target velocity depending on the current gap between the \cv and lead vehicle:   
\begin{equation}
v^\text{target} =
U+v^\text{catch}\times
\min(\max(\tfrac{\Delta x-g_{l}}{g_{u}-g_{l}},0),1)\;,
\end{equation}
which is up to  $v^\text{catch}$ above $U$.  This allows the \catvehicle to drive faster than the average velocity and catch up to the lead vehicle, should it face a gap above the lower threshold $g_{l}$, while at lower gaps the target velocity reduces to the average $U$. At gaps above the upper gap limit $g_l$, \cv vehicle should close the gap by traveling $v^\text{catch}$ above $U$. 
 
The commanded velocity sent to the low level \cv controller is updated via the rule
\begin{equation}
v^\text{cmd}_{j+1} =
\beta_j(\alpha_j v^\text{target}_j + (1-\alpha_j)v^\text{lead}_j)
+ (1-\beta_j)v^\text{cmd}_j\;,
\label{eq:cmdrule}
\end{equation}
where the subscript $j$ denotes the time step. This rule~\eqref{eq:cmdrule} chooses the new commanded velocity as a weighted average of the prior commanded velocity, the target velocity, and the lead vehicle's velocity. The weights $\alpha_j$ and $\beta_j$ depend on the gap as follows: $\beta = 1-\frac{1}{2}\alpha$, and 
\begin{equation}
\alpha = \min(\max(\tfrac{\Delta x-\Delta x^\text{s}}{\gamma},0),1)\;.
\label{eq:definealpha}
\end{equation}
In \eqref{eq:definealpha}, the distance $\Delta x^\text{s}$ is a safety distance. We have $\alpha = 0$ if $\Delta x\le\Delta x^\text{s}$ and $\alpha = 1$ if $\Delta x\ge\Delta x^\text{s}+\gamma$, meaning that for relatively short gaps, only the lead vehicle's velocity matters, while for relatively large gaps, only the target velocity is averaged with the commanded velocity. The parameter $\gamma$ controls the the rate at which $\alpha$ transitions from 0 to 1, and is set to $\gamma=2\text{ m}$ in the current implementation. This means that when the gap is short, the \catvehicle has the same speed of the lead vehicle, while when the gap is larger the \catvehicle speed tends towards the target vehicle, which allows the  \cv to reduce the gap with the lead vehicle.   
The parameter $\beta$ determines how rapidly the controller adjusts to new situations (with more rapid adjustments occurring in more safety-critical situations). At its core, this is a PI controller, but with a saturation at small gaps (for safety purposes), and a saturation at large gaps (so that the \cv closes gaps).

The model parameters for both controllers were determined via testing in a simulation environment (with a data-fitted human-driver model), as well as via car-following field tests (before the actual experiments). As a result we set the lower gap limit $g_l= 7$ m, the upper gap limit as $g_u=30$ m and $v^\text{catch}=1 \text{ m/s}$. The safety distance is implemented as $\Delta x^\text{s} = \max( 2\text{ s}\times \Delta v , 4\text { m})$. The term $2\text{ s}\times \Delta v$ represents the recommended following distance according to the 2 seconds rule, with a lower bound of $4\text{ m}$.

\subsection{Low level vehicle controls}
The commanded velocity produced by the two controllers described above is translated to the actual vehicle controls (i.e., gas and brake signals) via a multi-mode controller. The use of a multi-mode controller permits different gains to be used for acceleration or for braking, to enable faster braking when needed while avoiding chattering at steady-state velocities. 

Each mode is a PID controller, with gains determined through system identification of the \cv at constant velocities representative of those recorded in the experiment, using a similar structure as in \cite{6882826}. The \cv plant is simplified as a first-order model based on constant accelerator inputs. The controller's design is thus:
\begin{equation*}
a_{j+1} =
\begin{cases}
h_1(v_j,v^\text{cmd}_j) & \mathrm{if~~} v^\text{cmd}_j - v_j > - 0.25\frac{\text{m}}{\text{s}} \\
h_2(v_j,v^\text{cmd}_j) & \mathrm{if~~} v^\text{cmd}_j - v_j \leq - 0.25\frac{\text{m}}{\text{s}}\\
0 & \mathrm{otherwise}\;,
\end{cases}
\end{equation*}
in which $a_{j+1}\in [-100,100]$ represents the next commanded ``acceleration'' value, where 100 is the maximum depression of the accelerator, and -100 is the maximum depression of the brake. Moreover, $v_j$ is the current speed of the \cv, and $v_j^\text{cmd}$ is the desired speed. When $a < 0$ the brake is depressed, and when $a > 0$ the accelerator is depressed. Controller $h_1$ is designed to accelerate to the desired reference speed and maintain that desired speed primarily through control of the vehicle's accelerator, and controller $h_2$ is designed to effect more rapid speed reduction via the brake. Thus when the desired speed is less than $0.25 \frac{\text{m}}{\text{s}}$ of the current speed, the brake is used, and otherwise the accelerator is used to control speed (as in normal driving when releasing the accelerator reduces speed). These controllers are provided sampled data at 20 Hz and are permitted to send new updates to the \cv at 20Hz.

The performance characteristics of each controller are provided for a change of input as a step function of 1 m/s (-1 m/s for braking). The controller $h_1$ has a rise time of approximately 1.6 s with an overshoot of 5\% and a settling time of approximately 5.52 s. The controller $h_2$ has a rise time of approximately 0.8 s, with an 11\% overshoot and settling time of approximately 1.94 s. 

Given the dynamics of the \cv, a tradeoff must be performed on the comfort of the ride and the physical dynamics for a change in reference speed. The switching nature of the controller provides robustness to noise in sampled speed, since the accelerator is primarily used to control speed at steady state. Finally, the PID controllers are reset at 0 velocity, and standard approaches for windup avoidance are used to prevent unsafe acceleration \cite{astrom2005advancedpid}. 

\subsection{Human driver controller}
\label{sec:human_controller}
The driver who implements human control of the CAT Vehicle (coauthor M. Bunting) is instructed to attempt to maintain a desired velocity, but to slow down to avoid collision with the vehicle ahead. This is similar to the control law used in Experiment A, with the notable exception that the desired velocity is given in miles/hour (the primary readout of the speedometer in the \cv). The driver received training from the University of Arizona on safe driving of high-occupancy vehicles, and had extensive practice to drive in this way, before the actual experiment is performed.

\section{Experimental results: Dampening traffic waves with a single vehicle}
\label{sec:analysis_of_experimental_data}

The experimental results are presented in this section. To be able to effectively compare the results of the experiments, it is important to define metrics, which are consistent across the experiments. To this end, we present the metrics used to describe the traffic flow in Section~\ref{sec:metrics}.  With the metrics fully defined, we  present the results of the three experiments conducted in Section~\ref{sec:results}. 

\subsection{Definition and calculation of metrics}
\label{sec:metrics}
At each time step, we have the following data. From the image processing, we have the position $x$, the velocity $v$, and the acceleration $a$ of each vehicle. From the OBD-II sensors, we have the instantaneous fuel consumption $c$ of each vehicle. Let $f^i_j$ denote the sample of a quantity $f$, corresponding to vehicle $i$, at time $t_j$. A temporal average of a quantity over an interval $t\in [t_\text{start},t_\text{end}]$ is calculated as $\bar{f}^i = \frac{1}{m}\sum_{j=1}^m f^i_j$, where $f^i_1,\dots,f^i_m$ are the samples of vehicle $i$ in that time interval. Likewise, an average over all $n$ vehicles at an instant $t_j$ is given by $\bar{f}_j = \frac{1}{n}\sum_{i=1}^n f^i_j$, where $j$ denotes the time step. Finally, spatio-temporal averages are given by $\bar{f} = \frac{1}{mn}\sum_{i=1}^n \sum_{j=1}^m f^i_j$. The precise quantities of interest are defined next.


At each time instant we compute the spatially-averaged instantaneous velocity is computed by summing the velocity of each vehicle $i = 1, \cdots, n$ at a given time indexed by $j$, and dividing by the number of vehicles as: 

\begin{equation*}\label{eq:inst_vel}
    \bar{v}_j = \frac{1}{n}\sum_{i=1}^n v^i_j.
\end{equation*}

Over a given time interval with $m$ velocity samples per vehicle,  we compute the average (over all vehicles and over the time interval) as:

\begin{equation*}
    \bar{v} = \frac{1}{mn}\sum_{j=1}^m \sum_{i=1}^n v^i_j.
\end{equation*}

Similarly, we compute the velocity standard deviation of all vehicles and over the interval as:
\begin{equation*}
    \sigma~= \left(\frac{1}{mn-1}\sum_{j=1}^m \sum_{i=1}^n (v^i_j-\bar{v})^2\right)^\frac{1}{2}.
\end{equation*}

Given the fuel consumption of vehicle $i$ at time $t_j$, the average (over all vehicles and over a time interval)  consumption is computed as:
\begin{equation*}
    \bar{c} = \frac{1}{mn}\sum_{j=1}^m \sum_{i=1}^n c^i_j.
\end{equation*}

The throughput of traffic is computed as the product of the average velocity and the density (obtained from the number of vehicles and the length of the track $L=260$ m), and is given as as:
\begin{equation*}
   q = \frac{n}{L}\bar{v}.
\end{equation*}

We also quantify braking events. Given a threshold deceleration $\tau$, a brake event (deceleration peak) is defined as a contiguous region in time when $-a^i_j > \tau$ that has the additional property that the signal $-a^j_i$ must drop by more than $\tau$ on either side of a peak. This deceleration peak count is encoded in the function $\rho_{\tau}$, which takes the signal $a^i_1,\dots,a^i_m$ as an input, and outputs the number of peaks in that interval for vehicle $i$. The final calculation of $\kappa$, the rate of braking events, normalizes by the number of vehicles and total distance traveled (in kilometers). The threshold $\tau$ is chosen as the average standard deviation of deceleration, taken over all vehicles in the uncontrolled interval when waves are active. This is computed as:
\begin{equation*}
    \kappa = \frac{1}{n} \sum^n_{i=1} \frac{1}{x^i_m-x^i_1} \rho_{\tau}(a^i_1,\dots,a^i_m),
\end{equation*}
where $\tau$ is always calculated over the interval when waves are active. The interval is determined based on the standard deviation of the velocity as shown in  Appendix~\ref{sec:onset}.

\begin{table*}[t]
\centering\begin{tabular}{c c c c c c}
\toprule
\multirow{3}{0.115\textwidth}{\centering\textbf{Interval}} 
& \multirow{3}{0.115\textwidth}{\centering\textbf{Time (s)}}
& \multirow{3}{0.115\textwidth}{\centering\textbf{Velocity st. dev. (m/s)}}
& \multirow{3}{0.115\textwidth}{\centering\textbf{Fuel consumption ($\boldsymbol{\ell}$/100km)}}
& \multirow{3}{0.14\textwidth}{\centering\textbf{Braking (events/ vehicle/km)}}
& \multirow{3}{0.14\textwidth}{\centering\textbf{Throughput (veh/hr)}}\\ 
& & & & & \\
& & & & & \\
\midrule
Exp. start
 &  0 
 & 1.87 
 & 18.8 
 & 1.66 
 & 1809 
\\
Waves
 start
 & 79 
 & 3.31 
 & 24.6 
 & 8.58 
 & 1827 
\\
Autonomy
 6.50m/s
 & 126 
 & 1.69 
 & 18.0 
 & 3.45 
 & 1780 
\\
Autonomy
 7.00m/s
 & 222 
 & 0.67 
 & 15.0 
 & 0.21 
 & 1915 
\\
Autonomy
 7.50m/s
 & 292 
 & 0.64 
 & 14.1 
 & 0.12 
 & 2085 
\\
Autonomy
 8.00m/s
 & 347 
 & 1.56 
 & 17.7 
 & 2.50 
 & 1952 
\\
Autonomy
 7.50m/s
 & 415 
 & 1.14 
 & 16.7 
 & 0.31 
 & 1938 
\\
Disable
 Autonomy
 & 463 
 & 1.44 
 & 17.4 
 & 2.95 
 & 2133 
\\
Exp.
 end& 567& - & - & - & -\\
\bottomrule
\end{tabular}
\caption{Summary metrics over all vehicles by interval with corresponding start time, for Experiment A.}
\label{tab:metrics_table_testA}
\end{table*}

\subsection{Experimental results}\label{sec:results}
Experiment A contains 21 vehicles, including the \cv. The \cv is initially under human control, and the first traffic wave is observed 79 seconds into the experiment. The \emph{FollowerStopper} wave-dampening controller is activated 126 seconds into the experiment, and set with a desired velocity of $U=6.50$ m/s. Over the next several minutes, the desired velocity is varied step by step to test the dependence of the traffic conditions on the set point. It is changed to 7.00 m/s (222 seconds into the experiment), 7.50 m/s (292 seconds into the experiment), and finally 8.00 m/s at 347 seconds into the experiment. At 415 seconds, the desired velocity is reduced to 7.50 m/s, where it remains for 48 seconds. 
At 463 seconds into the experiment, 
the human driver resumes control of the \cv speed. The experiment is ended at 567 seconds. 

Experiment B also involves 21 vehicles and follows a similar design as Experiment A. The main difference is that in Experiment B, after the wave initially appears, a trained human driver implements the control strategy described in Experiment A but without the aid of automation. The human-executed control strategy is to maintain a desired velocity, calculated by an external observer as the average velocity of the previous lap, without colliding with the vehicle ahead. The \cv is always under human control, but the driver switches from initially following the instructions given to all human drivers to mimicking the control strategy in Experiment A after a wave appears. A traffic wave is first observed 55 seconds from the start of Experiment B, and the active control to dampen the wave begins after 112 seconds at a desired velocity of 6.25 m/s (14 mph in the units displayed in the \cv dashboard) with the command to ``drive with an average speed of 14 miles per hour, unless safety requires slower speeds.''
After 202 seconds, the \cv operator is instructed to increase the desired speed ``to 16 miles per hour,'' (7.15 m/s) 
which is maintained for 98 seconds before reverting to typical human driving behavior. The experiment is ended after 409 seconds. 

Experiment C is conducted with 22 vehicles. Note that one vehicle was added for this experiment, to demonstrate that instabilities and wave damping are not specific to having exactly 21 vehicles on the ring. At 161 seconds into the experiment, a traffic wave appears. At 218 seconds, the \emph{PI controller with saturation} 
wave damping controller is activated, and remains active until the end of the experiment at 413 seconds. Because the controller directly determines the desired velocity as part of the control algorithm, there are no external parameters that are changed during the experiment. 

After conclusion of the experiments, the data gathered through a $360^{\circ}$~camera placed at the center of track, in-vehicle devices, and \cv control computers are analyzed to characterize the performance of each experiment with respect to dampening traffic waves.

Displacement data describe each vehicle's distance traveled from their initial position on the ring at the beginning of the experiment. These data are extracted from the central camera using standard computer vision techniques 
to track vehicle positions along the ring. Velocity data are derived through discrete differentiation of displacement data (likewise, acceleration from velocity data)~\cite{WuTRB2017}, which are validated against the \cv control computers and the in-vehicle devices. The full data set and code used to generate the plots can be found in \cite{sternetal2016}.

Notable events in each experiment define time intervals over which to evaluate the state of traffic. In each analysis we quantify the traffic state according to (a) the standard deviation of traffic velocity; (b) the fuel consumption~\cite{Larrick1593}; (c) the rate of excessive braking events; and (d) the traffic throughput. 
A detailed description of how these calculations are performed, as well as the specific definitions of common metrics of fuel consumption, velocity (average and standard deviation), and excessive braking are available in Section~\ref{sec:metrics}.

\begin{figure}

        \centering
        
        \begin{subfigure}[b]{\textwidth}\centering
        \includegraphics[width=0.8\textwidth]{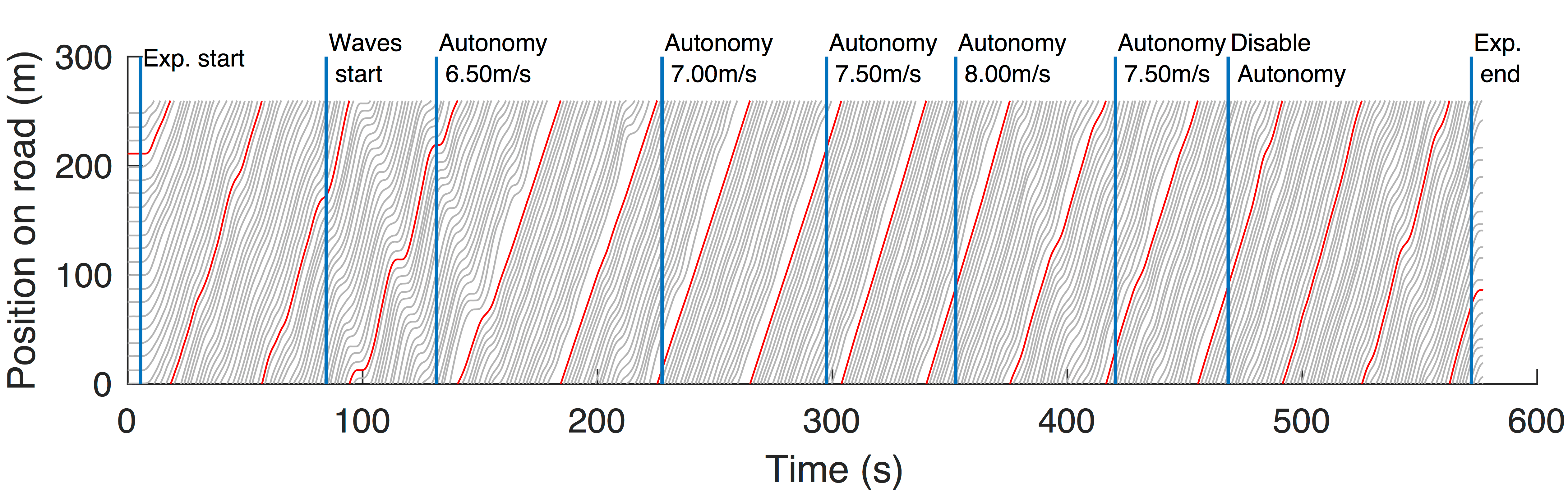}
        \caption{Trajectories of all vehicles in Experiment A, \cv shown in red.}
        \label{fig:test05_traj}
        \end{subfigure}\\

        \begin{subfigure}[b]{\textwidth}\centering
        \includegraphics[width=0.8\textwidth]{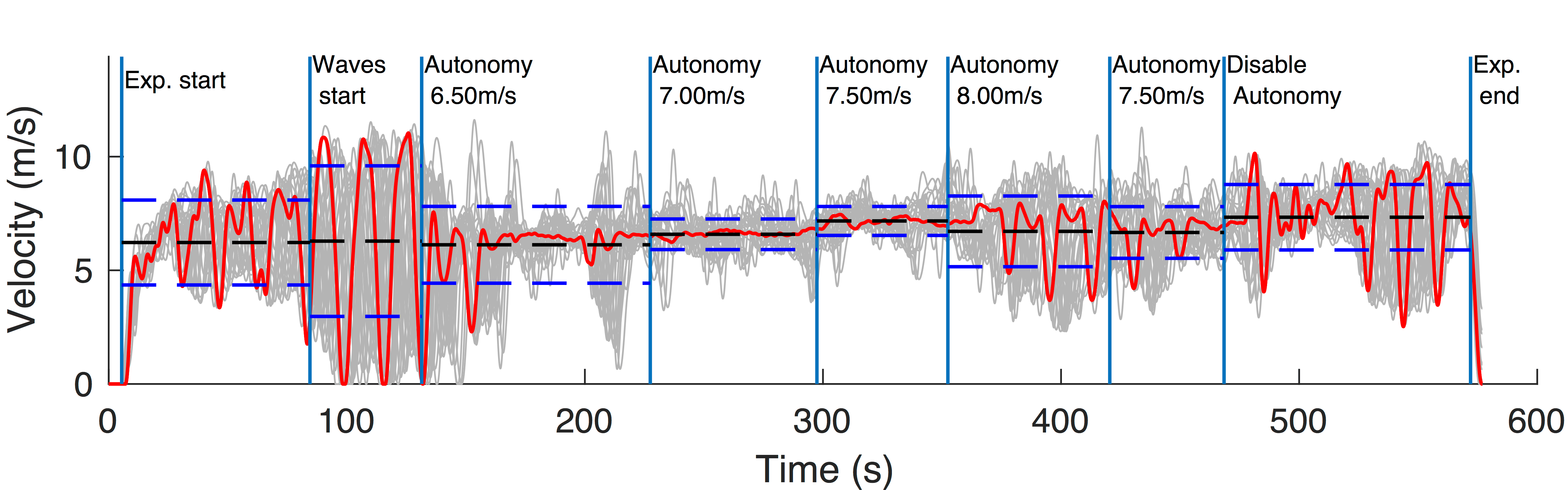}
        \caption{Velocity profiles of all vehicles (gray) and the \cv (red) in Experiment A. Horizontal blue dashed lines are one standard deviation above and below the mean speed of traffic in the interval.}
        \label{fig:test05_stDev}
        \end{subfigure}\\

        \caption{Trajectories and standard deviation in velocity for Experiment A.} 
        \label{fig:test05}
\end{figure}

In Experiment A, the traffic state is quantified over eight time intervals throughout the experiment. In each interval, the fuel consumption, velocity standard deviation, excessive braking, and throughput
are reported in Table~\ref{tab:metrics_table_testA}. The position of each vehicle's center over time is shown in Figure~\ref{fig:test05_traj}, with the \cv is shown in red and all other vehicles are shown in gray. In Figure~\ref{fig:test05_stDev}, the velocity profile of the \cv is shown in red, and all other vehicle velocities are plotted in gray. For each interval, the black dashed line denotes the average velocity of traffic over that interval, and the blue dashed lines denote the average velocity plus/minus one standard deviation.

When the \cv is initially under human control, a wave begins to appear. After the controller is activated 
the wave is noticeably affected, but not fully removed, through the interval when the \cv control is activated at a desired velocity of 6.50 m/s.
The wave-dampening effect of the controller is observed in the velocity profiles (Figure~\ref{fig:test05_stDev}), which exhibit a lower magnitude of oscillation from the mean after control begins.
When the \cv desired speed set in the \emph{FollowerStopper} is increased to 7.00 m/s 222 seconds into the experiment, further wave dampening is observed, and at a desired speed of 7.50 m/s, the best performance of the controller is achieved. Compared to the initial period where a wave was present under human control, the velocity standard deviation is reduced by 80.8\%, the fuel consumption is reduced by 42.5\%, and the excessive braking events are reduced from 8.58 events per vehicle per kilometer to 0.12 events per vehicle per kilometer. Because the average velocity of traffic on the ring is also increased, the throughput on the roadway increases by 14.1\%.

At 7.50 m/s, the \cv's speed matches the average traffic speed almost precisely, and consequently it does not need to slow down. However, at 347 seconds, the \cv desired velocity is increased further to 8.00 m/s and the \cv is, on average, faster than the flow of traffic---which inevitably induces a wave again.
The reappearance of a wave has the effect of increasing fuel consumption relative to the slower desired velocities, but still represents a benefit of the control compared to the uncontrolled traffic. At 415 seconds the \cv's desired velocity is reduced to 7.50 m/s, and the wave is once again dampened. The \emph{FollowerStopper} controller is deactivated after 463 seconds, and the traffic wave reappears. 

The control of the \cv has the impact of reducing the total fuel consumption of the traffic, as shown in Table~\ref{tab:metrics_table_testA}. The lowest fuel consumption of 14.13 $\ell$/100km is observed when the \emph{FollowerStopper} is operated with the set point of 7.50 m/s. This is also the lowest fuel consumption amongst all experiments conducted. A video of of Experiment A is provided in Movie S1 in the supplementary materials. 

\begin{figure}
        \centering
        
        \begin{subfigure}[b]{\textwidth}
        \centering
        \includegraphics[width=0.75\textwidth]{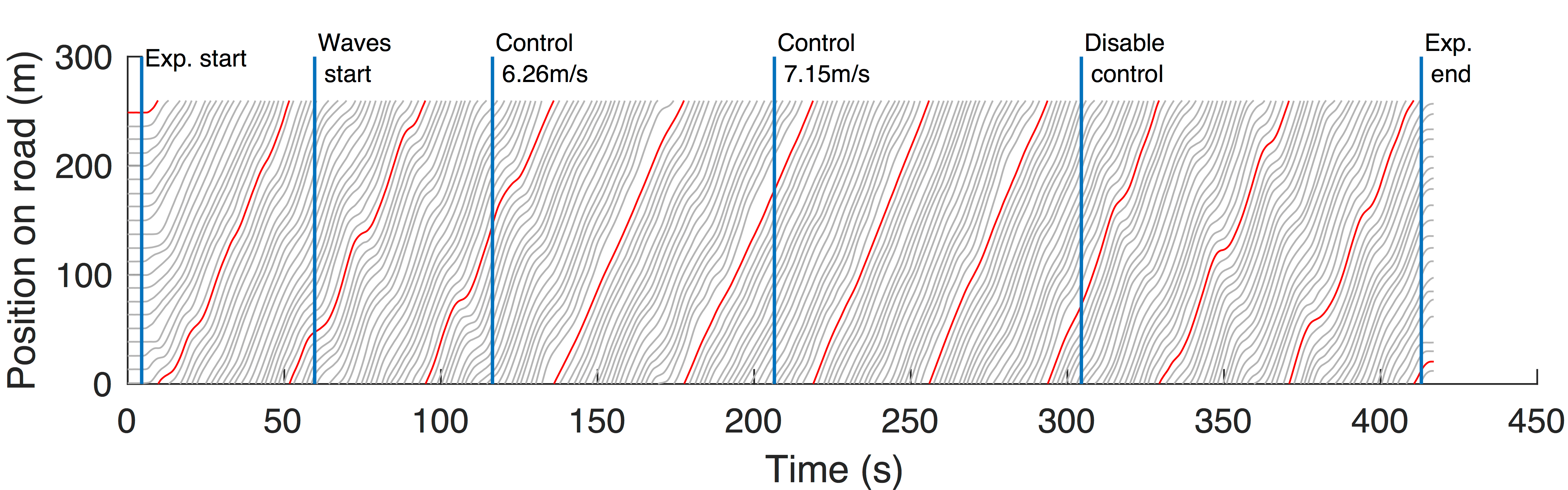}
        \caption{Trajectories of all vehicles in Experiment B, \cv shown in red.}
        \label{fig:test02_traj}
        \end{subfigure}\\
        
        \begin{subfigure}[b]{\textwidth}
        \centering
        \includegraphics[width=0.75\textwidth]{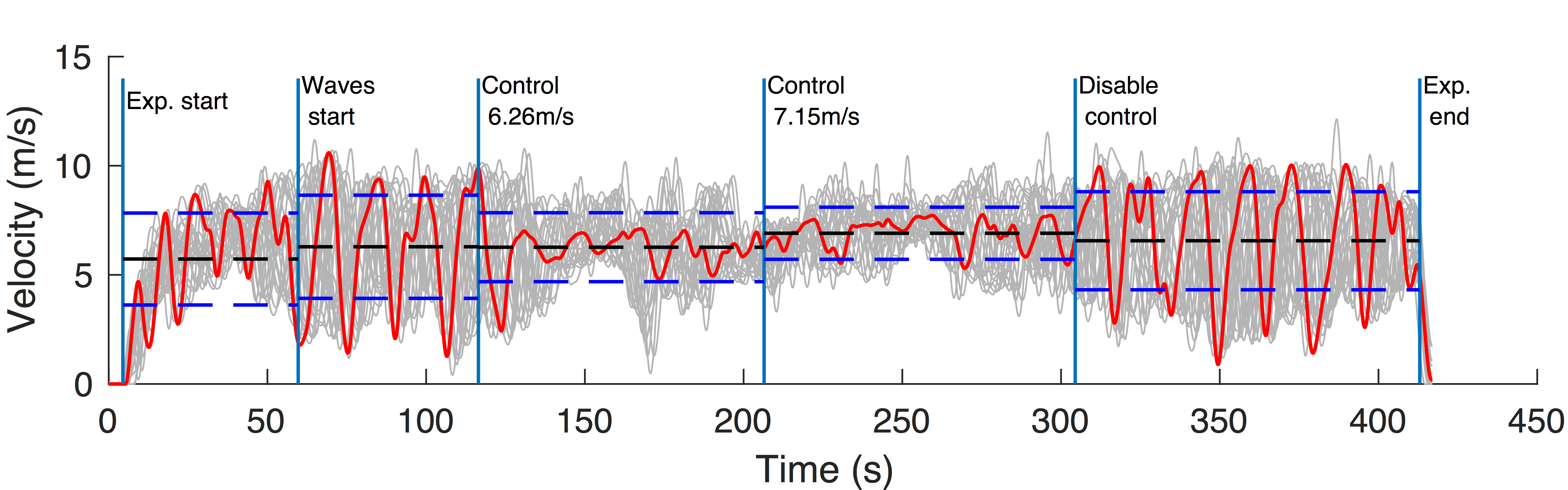}
        \caption{Velocity profiles of all vehicles (gray) and the \cv (red) in Experiment B. Horizontal blue dashed lines are one standard deviation above and below the mean speed of traffic in the interval.}
        \label{fig:test02_stDev}
        \end{subfigure}\\
        
        \caption{Trajectories and standard deviation in velocity for Experiment B.
}\label{fig:test02}
\end{figure}

In Experiment B, the \cv operator initially drives according to the same instructions as the other vehicle operators and a traffic wave appears at 55 seconds. At 112 seconds into the experiment, the \cv operator begins to drive at a desired velocity of 6.26 m/s without colliding with the vehicle in front. Later the desired velocity is increased before returning to the gap closing instructions 
followed by all other human drivers. The trajectories are shown in Figure~\ref{fig:test02} and the traffic state is quantified in each interval in Table~\ref{tab:metrics_table_testB}. 

\begin{table}
\centering\begin{tabular}{c c c c c c}
\toprule
\multirow{3}{0.12\textwidth}{\centering\textbf{Interval}} 
& \multirow{3}{0.12\textwidth}{\centering\textbf{Time (s)}}
& \multirow{3}{0.12\textwidth}{\centering\textbf{Velocity st. dev. (m/s)}}
& \multirow{3}{0.12\textwidth}{\centering\textbf{Fuel consumption ($\boldsymbol{\ell}$/100km)}}
& \multirow{3}{0.15\textwidth}{\centering\textbf{Braking (events/ vehicle/km)}}
& \multirow{3}{0.15\textwidth}{\centering\textbf{Throughput (veh/hr)}}\\ 
& & & & & \\
& & & & & \\
\midrule
Exp. start
 &  0 
 & 2.11 
 & 20.6 
 & 3.88 
 & 1665 
\\
Waves
 start
 & 55 
 & 2.36 
 & 21.8 
 & 9.50 
 & 1828 
\\
Control
 6.26m/s
 & 112 
 & 1.58 
 & 17.9 
 & 4.22 
 & 1822 
\\
Control
 7.15m/s
 & 202 
 & 1.19 
 & 17.0 
 & 2.27 
 & 2008 
\\
Disable
 control
 & 300 
 & 2.25 
 & 21.6 
 & 9.43 
 & 1908 
\\
Exp.
 end& 409& - & - & - & -\\
\bottomrule
\end{tabular}
\caption{Summary metrics over all vehicles by interval with corresponding start time, for Experiment B.}
\label{tab:metrics_table_testB}
\end{table}

The dampening effect of the human-implemented controller is quantified by the standard deviation of the velocities, which is reduced by 49.5\% when the 6.26 m/s control is active compared to when it is not. Similarly, excessive braking is reduced by 76.2\% when control is applied.

\begin{table}\centering\begin{tabular}{c  c c c  c c c  c c c  c c c}
\toprule
\multicolumn{1}{c|}{\centering\textbf{Exp.}} &
\multicolumn{3}{|c|}{\centering\textbf{Velocity st.}} &
\multicolumn{3}{|c|}{\centering\textbf{Fuel consump-}} &
\multicolumn{3}{|c|} {\centering\textbf{Braking}}&
\multicolumn{3}{|c} {\centering\textbf{Throughput}}\\
\multicolumn{1}{c|}{~} &
\multicolumn{3}{|c|}{\centering\textbf{dev. (m/s)}} &
\multicolumn{3}{|c|}{\centering\textbf{tion ($\boldsymbol{\ell}$/100km)}} &
\multicolumn{3}{|c|}{\centering\textbf{(events/veh/km)}} &
\multicolumn{3}{|c}{\centering\textbf{(veh/hr)}}\\
\multicolumn{1}{c|}{~} &
\multicolumn{1}{|c}{WS} &
\multicolumn{1}{c}{CA} &
\multicolumn{1}{c|}{\%} &
\multicolumn{1}{|c}{WS} &
\multicolumn{1}{c}{CA} &
\multicolumn{1}{c|}{\%} &
\multicolumn{1}{|c}{WS} &
\multicolumn{1}{c}{CA} &
\multicolumn{1}{c|}{\%} &
\multicolumn{1}{|c}{WS} &
\multicolumn{1}{c}{CA} &
\multicolumn{1}{c}{\%} \\
\midrule
A
 & 3.31 
 & 0.64 
 & -80.8 
 & 24.6 
 & 14.1 
 & -42.5 
 & 8.58 
 & 0.12 
 & -98.6 
 & 1827 
 & 2085 
 & +14.1 
\\
B
 & 2.36 
 & 1.19 
 & -49.5 
 & 21.8 
 & 17.0 
 & -22.1 
 & 9.50 
 & 2.27 
 & -76.2 
 & 1828 
 & 2008 
 & +9.8 
\\
C
 & 3.85 
 & 1.74 
 & -54.7 
 & 29.0 
 & 20.9 
 & -28.1 
 & 9.66 
 & 2.47 
 & -74.4 
 & 1755 
 & 1711 
 & -2.5 
\\
\bottomrule
\end{tabular}

\caption{Summary metrics of the flow in each experiment under the first interval when the traffic wave starts (WS) without wave dampening control, and under the best interval when control is active (CA). The percent change from WS to CA in each experiment is also reported.}
\label{tab:experiment_comparison_table}
\end{table}

The desired speed given to the \cv driver also influences the reduction in the velocity variability. When the desired speed of the \cv is increased to 16 mph (7.15 m/s), the standard deviation reaches the minimum values for the experiment. The throughput is also higher during this period compared to when the traffic is uncontrolled (2008 veh/hr vs.~1828 veh/hr). Once the \cv returns to a typical gap closing behavior (i.e., no longer under human control to maintain a desired speed), the traffic wave reappears, and the velocity standard deviation increases.

When the \cv is under wave dampening control by a human, the control reduces the overall fuel consumption when compared to the uncontrolled case (Table~\ref{tab:metrics_table_testB}), where average fuel consumption ($\ell$/100km) is recorded for each period of the experiment. The result is a decrease in fuel consumption when the \cv begins to dampen the traffic wave (a decrease from 21.8 $\ell$/100km to 17.9 $\ell$/100km, (17.9\%)). A further decrease in fuel consumption  (to 17.0 $\ell$/100km, a decrease of 22.0\% compared to when a wave is present initially) is observed when the desired velocity is increased from 14 mph (6.25 m/s) to 16 mph (7.15 m/s). Finally, when the \cv returns to human-driven behavior and stops actively dampening the traffic wave, fuel consumption increases again and returns to a level similar to the pre-control fuel consumption  (21.8 $\ell$/100km) before control starts compared to (21.6 $\ell$/100km) after control ends and the traffic wave reappears).

Finally, in Experiment~C, an additional vehicle is added to the track, bringing the total number of vehicles to 22. As in other experiments, the \cv begins the experiment in human control. The traffic is relatively smooth until the first strong wave occurs at 161 seconds. Comparing the traffic conditions under no control before the strong wave appears to when it is observed, the traffic wave results in a 
37.7\% increase in the average velocity standard deviation, a 63.8\% 
increase in fuel consumption, and a 18.8\% 
reduction in throughput.


\begin{table}\centering\begin{tabular}{c c c c c c}
\toprule
\multirow{3}{0.12\textwidth}{\centering\textbf{Interval}} 
& \multirow{3}{0.12\textwidth}{\centering\textbf{Time (s)}}
& \multirow{3}{0.12\textwidth}{\centering\textbf{Velocity st. dev. (m/s)}}
& \multirow{3}{0.12\textwidth}{\centering\textbf{Fuel consumption ($\boldsymbol{\ell}$/100km)}}
& \multirow{3}{0.15\textwidth}{\centering\textbf{Braking (events/ vehicle/km)}}
& \multirow{3}{0.15\textwidth}{\centering\textbf{Throughput (veh/hr)}}\\ 
& & & & & \\
& & & & & \\
\midrule
Exp. start
 &  0 
 & 1.62 
 & 17.7 
 & 1.11 
 & 2160 
\\
Waves
 start
 & 161 
 & 3.85 
 & 29.0 
 & 9.66 
 & 1755 
\\
Autonomy
 & 218 
 & 1.74 
 & 20.9 
 & 2.47 
 & 1711 
\\
Exp.
 end& 413& - & - & - & -\\
\bottomrule
\end{tabular}
\caption{Summary metrics over all vehicles  by interval with corresponding start time, for Experiment C.}
\label{tab:metrics_table_testC}
\end{table}

At 218 seconds, the \emph{PI controller with saturation} wave damping traffic controller is activated, and the wave is substantially reduced. Compared to the interval when the wave is present, the controller results in a reduction of the speed variability 
(54.7\% reduction in standard deviation see Table~\ref{tab:metrics_table_testC}). It also reduces the fuel consumption by 
27.9\%, and the rate of excessive braking is reduced by
74.4\%. The throughput is also slightly reduced by
2.5\%.
Note that this controller only uses information directly measured by the \cv itself, and it does not require any external information. One consequence of this more local nature is that the wave dampening is not as perfect as in Experiment A. In particular, the controller slightly reduces the average velocity. Nevertheless, the velocity standard deviation, excessive braking, and fuel consumption are substantially reduced compared to when uncontrolled traffic waves are present.

\begin{figure}

        \centering
        
       \begin{subfigure}[b]{\textwidth}
       \centering
        \includegraphics[width=0.75\textwidth]{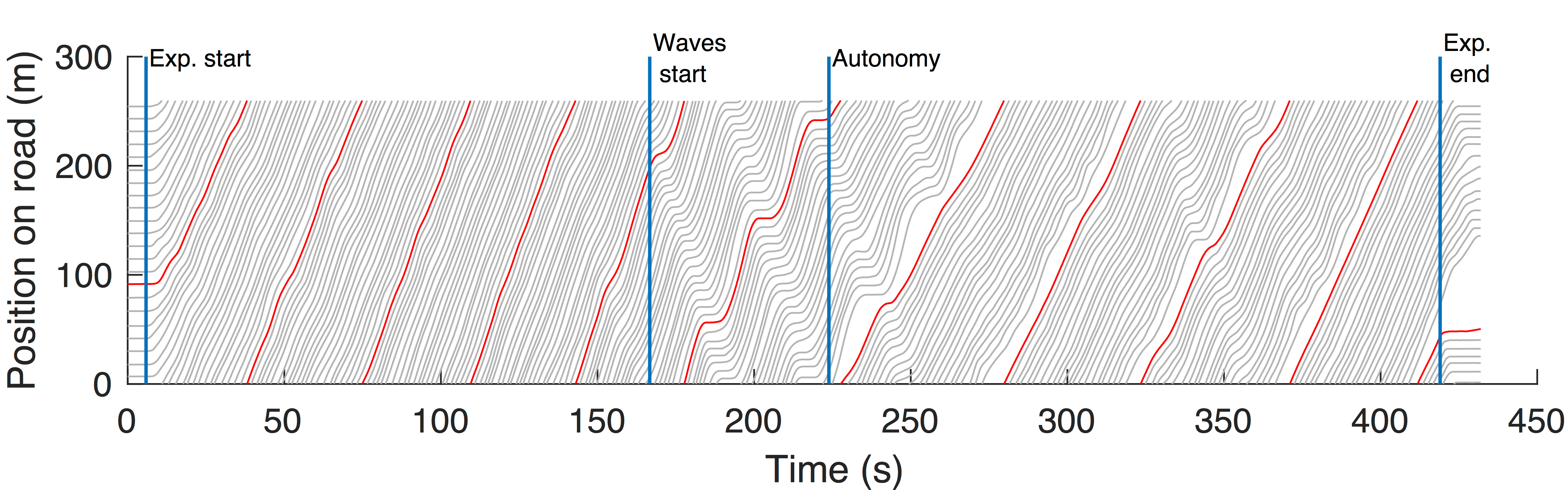}
        \caption{Trajectories of all vehicles in Experiment C, \cv shown in red.}
        \label{fig:test10_traj}
        \end{subfigure}\\        
        
        \begin{subfigure}[b]{\textwidth}
        \centering
        \includegraphics[width=0.75\textwidth]{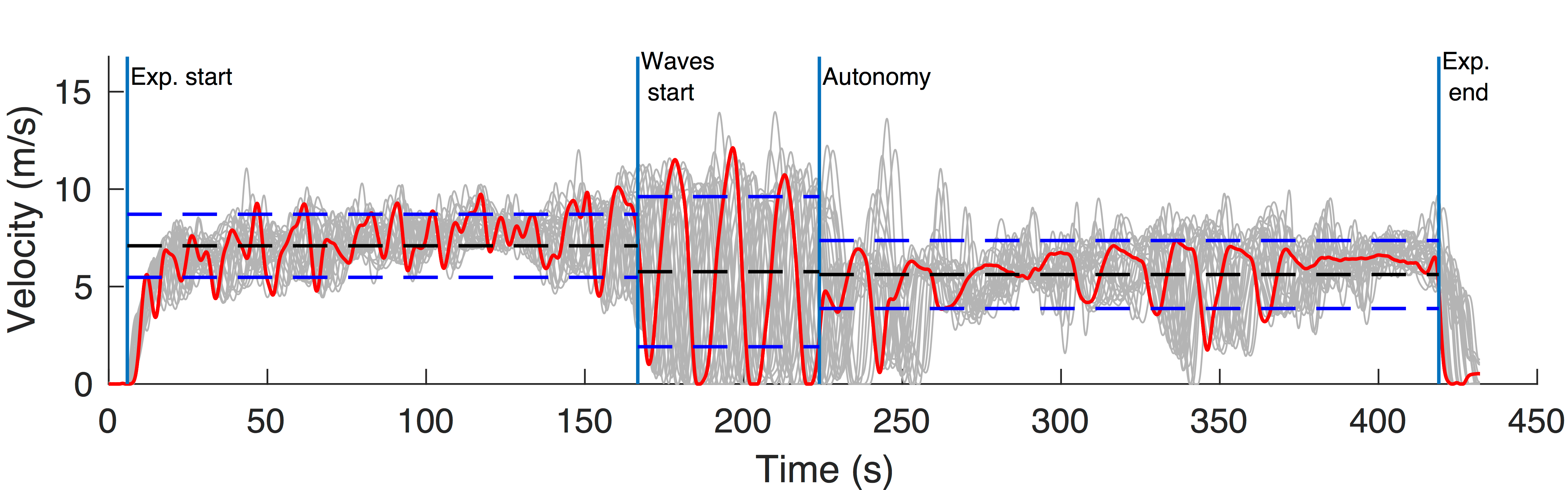}
        \caption{Velocity profiles of all vehicles (gray) and the \cv (red) in Experiment C. Horizontal blue dashed lines are one standard deviation above and below the mean speed of traffic in the interval.}
        \label{fig:test10_stDev}
        \end{subfigure}\\

        \caption{Trajectories and standard deviation in velocity for Experiment C.}\label{fig:test10}
\end{figure}

By examining the trajectories (Figure~\ref{fig:test10}), it is apparent that the control law is able to eliminate the initially present wave, but another wave is generated during the control period. The second wave is also damped by the control law, but its presence for a period of the active control accounts for the difference relative to the period where no waves occurred in the uncontrolled period at the start of the experiment.


To conclude, these experiments demonstrate that traffic flow control via low penetration rate automated vehicles is in fact possible. Moreover, the data collected quantify the benefits of conducting control via the AV. Specifically, under proper control, (a) the velocity standard deviation reduces noticeably; (b) the fuel consumption is reduced by a significant margin; (c) braking events are substantially reduced; and (d) in some experiments, even the average velocity (and thus the throughput) is increased (Table~\ref{tab:experiment_comparison_table}). 


\emph{Velocity standard deviation:} The velocity standard deviation is reduced in all experiments, ranging between 49.5\% for Experiment B, 54.7\% for Experiment C, and a high of 80.8\% for Experiment A.

\emph{Fuel consumption:} The fuel consumption in all experiments is reduced from when waves are present and under human control compared to when the autonomous car is active. The improvements include a reduction of 42.5\%, 22.1\%, and 28.1\% in Experiments A, B, and C, respectively.

\emph{Excessive braking:} The number of excessive braking events is also substantially reduced, from 8.58 to 9.66 excessive braking events/veh/km down to 2.47 
events/veh/km in the worst performing controller and nearly complete elimination (0.12 events/veh/km) in the best performing controller.

\emph{Throughput:} Changes in throughput for each experiment are +14.1\% for Experiment A, and +9.8\% for Experiment B, with a $-$2.5\% change in Experiment C.


\section{Conclusions}\label{sec:conclusions}
AVs can revolutionize the control of traffic flow. They offer the potential to shift from localized control measures, like ramp metering, and centralized ones, as variable speed limit gantries, to Lagrangian actuators immersed in the traffic stream. Strikingly, 
it is not necessary for all vehicles to be automated in order to benefit from mobile actuation. A single autonomous vehicle can control the flow of at least 20 human-controlled vehicles around it, with substantial reductions in velocity standard deviation, excessive braking, and fuel consumption.\\
Moreover, this study demonstrates that these benefits can be achieved via structurally very simple control strategies, based only on the AV's velocity, its spatial gap between the vehicle immediately in front, and some estimate of the average velocity of traffic flow. The \emph{PI with saturation} controller (Experiment~C) is a fully automatic control, while the \emph{FollowerStopper} (Experiment~A) and the human-implemented control (Experiment~B) have an external input (dependent on observed traffic conditions). This simple structure implies that a noticeable impact on congested traffic flow can in principle be achieved by means of adaptive cruise control systems that are already in place in certain new vehicles, and the use of intelligent infrastructure and/or connected vehicles to provide the required external inputs.

Most contemporary traffic control strategies (implemented in practice, and/or proposed in the literature) are based on centralized interventions, such as ramp-metering, variable speed limits, and traffic light controls. For example, a successful variable speed limit control may yield a 5\% increase in capacity \cite{nissan2011evaluation}. However, those traditional control approaches will always have limited effect on the traffic dynamics that emerge between the fixed control points. In contrast, the control of traffic flow via a sparse set of Lagrangian actuators (AVs or trained human drivers) enables new opportunities for control, with a direct positive effect on the dynamics of traffic flow, and without the need of a dedicated actuation infrastructure.

The presented ring experiments represent a stretch of single-lane roadway. However, the theory extends also to multi-lane freeways, on which lane changing can serve as an additional trigger of stop-and-go waves. The lane changes can also open up gaps in the vacated lane, which can serve to dampen waves in that lane. The central challenge in the multi-lane setting is to have controllers that dynamically dampen waves, but without leaving too large gaps, because large gaps may trigger additional lane changing, which may reduce the effectiveness of the strategy. The results of Experiment~A demonstrate that the controller does in fact not leave a large gap once the waves have been damped. To fully quantify the benefits of Lagrangian actuators on urban freeways, future multi-lane experiments are needed.

The control of complex multi-agent systems has impact beyond vehicular traffic flow, including 
coordinated robots~\cite{OlfatiSaberetal2007}, social networks~\cite{mesbahi2010graph}, animal swarming~\cite{Hoenicke2015,berdahl2013emergent},
and many other applications. 
However, in contrast to many applications in robotics or fleet control, the human agents play a crucial role in traffic flow dynamics. Moreover, in contrast to other human-in-the-loop cyber-physical systems, the automated controller and the human agents are spatially separated, and they do not work cooperatively. Rather, the AVs counteract the humans' tendency to produce unstable traffic situations. The results shown here imply that this concept is not a far future but instead could be, in principle, implemented with already existing technology.

\section*{Acknowledgements}
This material is based upon work supported by the National Science Foundation under Grant No. CNS-1446715 (B.P.), CNS-1446690 (B.S.), CNS-1446435 (J.S.), and CNS-1446702 (D.W.). The authors thank the University of Arizona Motor Pool in providing the vehicle fleet. They offer additional special thanks for the services of N. Emptage in carrying out the experiment logistics.


\section*{Appendix}


\subsection{Vehicle specifications}\label{sec:vehMake}

A full description of the year, make, model, length, and EPA rated fuel consumption of each vehicle used in the experiments is presented in Table \ref{tab:vehicle_summary}. For Experiments A and B, vehicles 1 through 21 are used. In Experiment C, vehicle 22 is also used.

\begin{table}
\centering\begin{tabular}{c c c c c c c}
\hline
\multirow{3}{0.07\textwidth}{\centering\textbf{Veh. Num.}}&
\multirow{3}{0.08\textwidth}{\centering\textbf{Year}}&
\multirow{3}{0.08\textwidth}{\centering\textbf{Make}}&
\multirow{3}{0.10\textwidth}{\centering\textbf{Model}}&
\multirow{3}{0.08\textwidth}{\centering\textbf{Length (m)}}&
\multirow{3}{0.155\textwidth}{\centering\textbf{Consumption City ($\boldsymbol{\ell}$/100km)}}&
\multirow{3}{0.155\textwidth}{\centering\textbf{Consumption Hwy. ($\boldsymbol{\ell}$/100km)}}\\
\\
\\

\hline
\hline
1 & 2013 & Chevrolet & Silverado & 5.22 & 15.67 & 10.68\\
2 & 2013 & Dodge & Grand Caravan & 5.15 & 13.83 & 9.47\\
3 & 2015 & Chevrolet & Malibu & 4.86 & 9.47 & 6.53\\
4 & 2012 & Chevrolet & Malibu & 4.87 & 10.69 & 7.13\\
5 & 2012 & Dodge & Grand Caravan & 5.15 & 13.83 & 9.47\\
6 & 2013 & Dodge & Grand Caravan & 5.15 & 13.83 & 9.41\\
7 & 2014 & Chevrolet & Malibu & 4.86 & 9.41 & 6.53\\
8 & 2016 & Chevrolet & Malibu & 4.92 & 8.71 & 6.37\\
9 & 2013 & Chevrolet & Impala & 5.09 & 13.07 & 7.84\\
10 & 2014 & Chevrolet & Malibu & 4.86 & 9.41 & 6.53\\
11 & 2016 & Chevrolet & Malibu Limited & 4.86 & 9.80 & 6.92\\
12 & 2015 & Chevrolet & Suburban & 5.69 & 14.71 & 10.22\\
13 & 2014 & Chevrolet & Silverado & 5.21 & 13.07 & 9.80\\
14 & 2014 & Dodge & Grand Caravan & 5.15 & 13.83 & 9.41\\
15 & 2012 & Chevrolet & Malibu & 4.87 & 10.70 & 7.13\\
16 & 2016 & Dodge & Grand Caravan & 5.15 & 13.83 & 9.41\\
17 & 2014 & Chevrolet & Malibu & 4.86 & 9.41 & 6.53\\
18 & 2012 & Chevrolet & Malibu & 4.87 & 10.70 & 7.13\\
19 & 2012 & Dodge & Grand Caravan & 5.15 & 13.83 & 9.41\\
20 & 2016 & Chevrolet & Suburban & 5.70 & 14.71 & 10.22\\
21 & 2009 & Ford & Escape Hybrid & 4.44 & 6.92 & 7.84\\
22 & 2012 & Dodge & Grand Caravan & 5.15 & 13.83 & 9.41\\
\hline
\hline

\end{tabular}
\caption{Vehicle properties of all vehicles used in the experiments.}
\label{tab:vehicle_summary}
\end{table}

\subsection{Identifying onset of oscillatory traffic behavior}
\label{sec:onset}
\begin{figure}

        \centering
        
        \begin{subfigure}[b]{\textwidth}
        \centering
        \includegraphics[width=0.7\columnwidth]{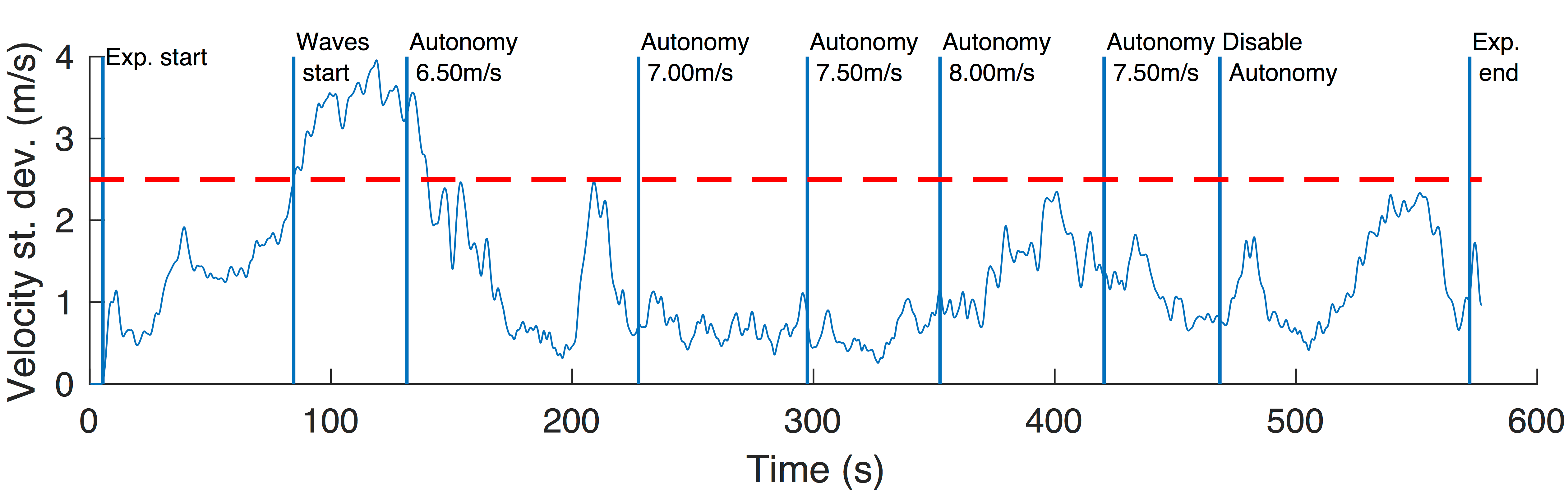}
        \caption{Instantaneous standard deviation of vehicle velocities for Experiment A with dashed line indicating threshold for traffic waves in flow.}
        \label{fig:inst_std_test05}
        \end{subfigure}\\
        
        \begin{subfigure}[b]{\textwidth}
        \centering
        \includegraphics[width=0.7\columnwidth]{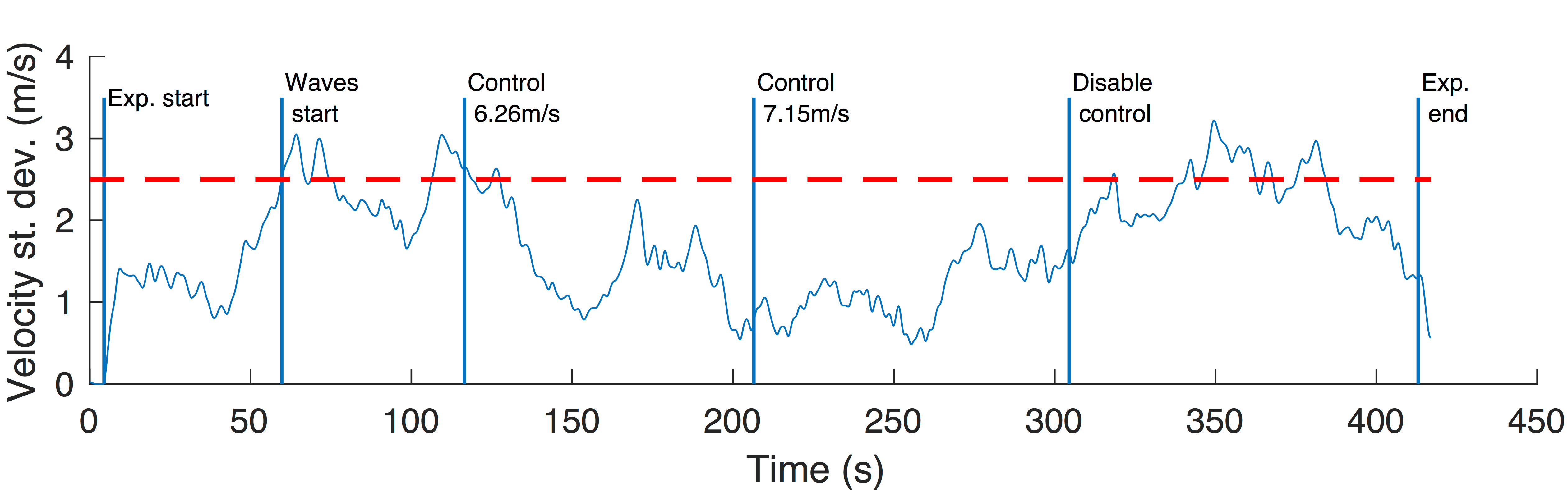}
        \caption{Instantaneous standard deviation of vehicle velocities for Experiment B with dashed line indicating threshold for traffic waves in flow.}
        \label{fig:inst_std_test02}
        \end{subfigure}\\
        
        \begin{subfigure}[b]{\textwidth}
        \centering
        \includegraphics[width=0.7\columnwidth]{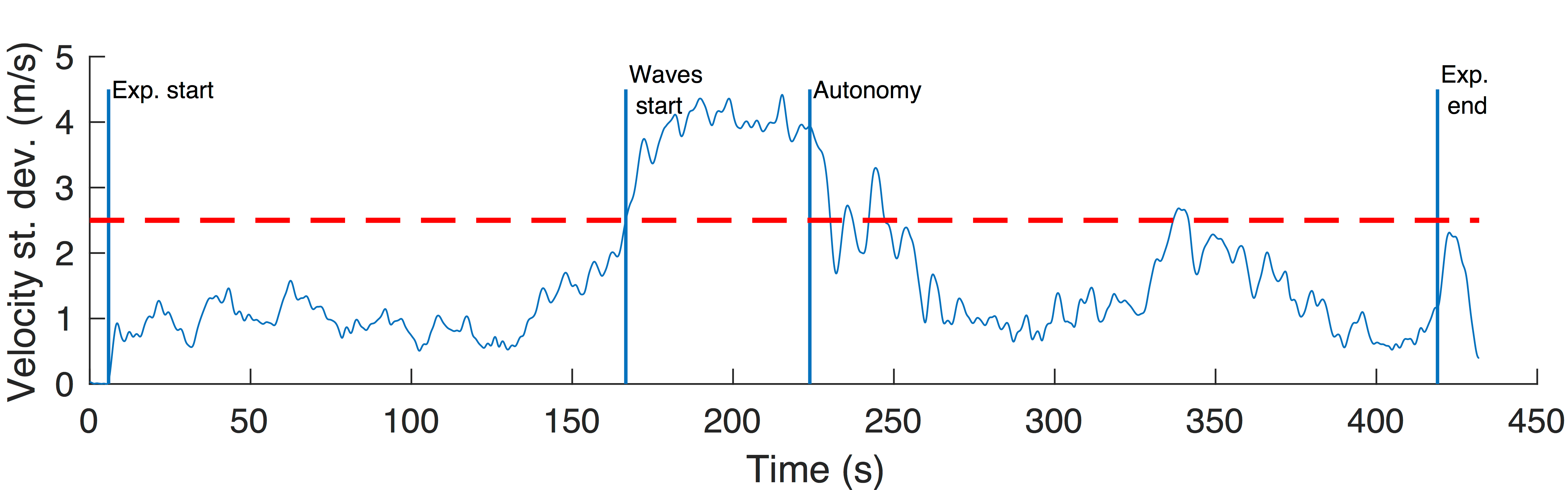}
        \caption{Instantaneous standard deviation of vehicle velocities for Experiment C with dashed line indicating threshold for traffic waves in flow.}
        \label{fig:inst_std_test10}
        \end{subfigure}\\

        \caption{Instantaneous velocity standard deviation for all three experiments showing onset of traffic wave.
}\label{fig:std_dev}
\end{figure}

We briefly describe the method used to identify the onset of oscillatory traffic behavior. The mean and standard deviation of the instantaneous vehicle velocities are computed at each timestep (i.e., for each timestep we compute the average of all vehicle speeds at that timestep, and then compute the sample standard deviation of all vehicle speeds at that timestep).   When the standard deviation of the instantaneous vehicle velocities exceeds 2.5 m/s, the traffic is considered to contain a traffic wave. This threshold is used to define the time at which waves first appear in the traffic experiments. Figures \ref{fig:inst_std_test05}, \ref{fig:inst_std_test02}, and \ref{fig:inst_std_test10} 
show a timeseries of the instantaneous velocity standard deviation for Experiments A, B, and  respectively  with the threshold plotted in a red dashed line.

\bibliography{refs}



\bibliographystyle{elsarticle-num}


\end{document}